\documentclass[aps,pra,reprint, superscriptaddress,showpacs]{revtex4-1}

\usepackage{amsmath,amssymb}
\usepackage{txfonts}
\usepackage[dvipdfmx]{graphicx}
\usepackage{subfigure}
\usepackage{url}
\begin{document}


\title{Data Search by a Laser Ising Machine with Gradual Pumping or Coupling}



\author{Kenta Takata}
\email[]{takata@nii.ac.jp}
\affiliation{National Institute of Informatics, Hitotsubashi 2-1-2, Chiyoda-ku, Tokyo 101-8403, Japan}
\affiliation{Department of Information and Communication Engineering, The University of Tokyo, Hongo 7-3-1, Bunkyo-ku, Tokyo 113-8654, Japan}

\author{Yoshihisa Yamamoto}
\email[]{yyamamoto@stanford.edu}
\affiliation{National Institute of Informatics, Hitotsubashi 2-1-2, Chiyoda-ku, Tokyo 101-8403, Japan}
\affiliation{Department of Information and Communication Engineering, The University of Tokyo, Hongo 7-3-1, Bunkyo-ku, Tokyo 113-8654, Japan}
\affiliation{E. L. Ginzton Laboratory, Stanford University, Stanford, California 94305, USA}



\date{\today}

\begin{abstract}
We study two operational schemes for a coherent Ising machine based on an injection-locked laser network.
These schemes gradually increase the pumping rate or the mutual coupling among the slave lasers.
We numerically simulate the two schemes against a data search problem implemented with the Ising model in cubic graphs without frustration.
We show the machine can achieve a better success probability and effective computational time to find a target/ground state with these gradual schemes
than those with the abrupt introduction of the mutual injection which has been studied previously.
The computational time simulated with typical parameters is almost constant up to the problem size $M = 200$
and turns into a nearly linear scale holding up to $M = 1000$.
\end{abstract}


\maketitle

\section{INTRODUCTION}
There are many combinatorial optimization problems intractable with modern digital computers and existing algorithms \cite{bib:NPC}.
Among these, NP problems can be solved in a polynomial time only with a hypothetical non-deterministic Turing machine.
In NP problems, it is examined if given instances meet certain conditions.
On the other hand, NP-hard problems basically require the most optimized solutions in various kinds of combinatorial optimization problems
and are as difficult as NP problems at least.
Both of them are believed to require the computational time growing exponentially with the problem size.
The problems that can be both NP and NP-hard are called NP-complete problems.
NP-complete problems can be mapped to each other with polynomial-time overheads.
Thus, solving an NP-complete problem efficiently is a holy grail for computer science and information technology \cite{bib:P_vs_NP}.

From the viewpoint of statistical physics, the ground state of the Ising Hamiltonian
${\cal H} = \sum _{i < j} J_{ij} \sigma _i \sigma _j + \sum _{i} \lambda _i \sigma _i$ is important to understand mysterious properties of spin glasses and magnetic disorders.
However, finding a ground state of the Ising Hamiltonian (Ising problem) of two dimensional lattices with external fields and that of three dimensional lattices are known to be NP-hard \cite{bib:Barahona}.
They can be reduced to NP-complete problems when we consider the decision problem if there is a state with an energy eigenvalue smaller than a prescribed value.
Also, other NP-complete problems such as MAX-CUT and graph partitioning can be easily mapped to anti-ferromagnetic Ising problems \cite{bib:Barahona2,bib:SP_Glass}.
Various computational schemes, such as simulated annealing \cite{bib:SA},
quantum annealing \cite{bib:QA_Ising},
and adiabatic quantum computation \cite{bib:AQC}
have been proposed and developed to solve such combinatorial optimization problems.
Especially, the adiabatic quantum computation has shown a good result for the NP-complete Exact Cover problem in small problem sizes \cite{bib:AQC_NPC}.
However, a further study has revealed that the problem size dependence of the computational time has a transition into an exponential scaling at a large problem size $M \sim 100$ \cite{bib:Young}.
Nevertheless, such an idea has triggered an experimental implementation of various quantum simulators and quantum annealers \cite{bib:QS_exp1,*bib:QS_exp2,*bib:QS_exp3,*bib:QS_exp4,*bib:QS_exp5}.
Also, the non-Hermitian quantum annealing scheme has recently achieved a high transition probability to get the target state in Grover's search problem \cite{bib:NLAQC}.

\begin{figure}[htbp]
	\includegraphics[clip, scale=0.3]{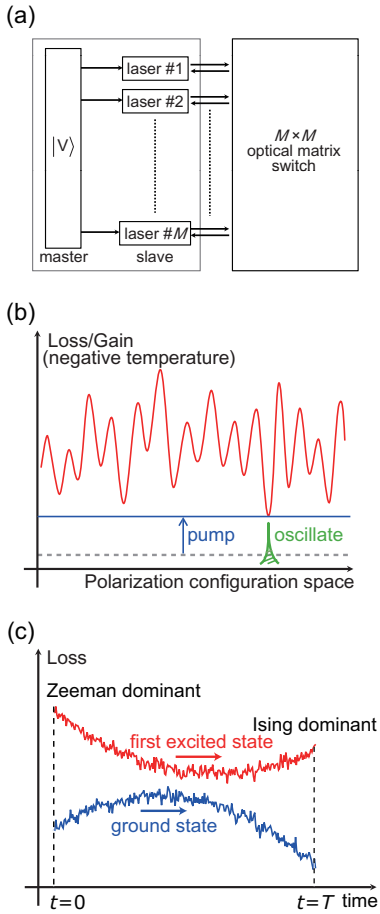}
	\caption{Schematic illustrations for (a) the coherent Ising machine with mutual optical coupling, (b) the loss and gain landscape in the gradual pumping scheme and (c) the gradual path between the initial and final states in the gradual coupling scheme.}\label{fig:laser_network}
\end{figure}

We have recently proposed a new computational machine \cite{bib:Utsunomiya_Takata_Yamamoto,bib:Takata_Utsunomiya_Yamamoto} for the Ising problem based on an injection-locked laser network [Fig. \ref{fig:laser_network}(a)].
In this coherent Ising machine, the normalized amplitude difference in circularly polarized modes of each slave laser is regarded as
an artificial spin with a continuous value.
Then, the Hamiltonian relaxed to a continuous function is embedded into the sum of the gain coefficients for all the slave lasers as a modulation induced by the mutual injection,
and gets dependent on the polarization configuration of the slave lasers ($\{|R \rangle, \, |L \rangle\}^M$).
The minimum gain coefficient means the minimum effective loss of photons and then the maximum number of photons in a laser.
Thus, when the mutual injection is introduced, the system spontaneously searches for a ground state of the mapped Hamiltonian with its bosonic nature.

In this paper, we study two operational schemes of the coherent Ising machine to improve its performance.
The first scheme is named ``gradual pumping'' (GP) scheme, where the pumping power into the slave lasers, i.e. the gain, is slowly increased [Fig. \ref{fig:laser_network}(b)].
In other well-known schemes such as simulated annealing and quantum annealing, the system temperature is gradually decreased and the quantum tunneling is gradually turned off, respectively.
In these cases, however, the system may be trapped in metastable excited states whose number increases exponentially in a hard instance of NP-complete/NP-hard problems.
In order to resolve such a dilemma, we map the energy landscape of a given problem to the net loss landscape of the network of gain media.
When we gradually increase the pumping power which is equivalent to the effective negative temperature of the inverted medium,
the first contact between the gain and polarization-dependent loss occurs at the ground state with the minimum loss.
Here, it is expected the system undergoes the phase transition from the initial state to the ground state in a way that the metastable excited states do not interfere with a computational process.
Note that the loss difference between the ground state and first excited state does not explicitly depend on the problem size $M$,
while the total spontaneous emission noise of the slave laser network is only proportional to $M$.

The second scheme is called ``gradual coupling'' (GC) scheme, which is inspired by the non-Hermitian annealing \cite{bib:NLAQC}.
In this case, each slave laser originally has fixed pumping level and only the vertically polarized injection signal ($ |V \rangle = \frac{1}{\sqrt{2}}(|R \rangle + |L \rangle) $).
Then, the mutual coupling between slave lasers is gradually increased [Fig. \ref{fig:laser_network}(c)].
Thus, the system is expected to show the phase transition from the master signal dominant state to the mutual injection dominant state
with the minimum total loss.
Note that this scheme is distinct from quantum annealing, because the coherent Ising machine utilizes macroscopic
coherent states of light in the open system with continual inflow and outflow of energy.
On the other hand, quantum entanglement between single particles cannot be exploited in this machine.

For a benchmark of them, we code a random $M$-bit data file $\{+1, \, -1 \}^M$ into the Ising (Mattis) Hamiltonian \cite{bib:Nishimori} in particular cubic graphs.
Here, the Hamiltonian reflecting the picked target state as a ground state is loaded to the system.
The ground state should be trivial, so the problem does not have any frustration and hard instances.
Instead, we expect consistent parameter dependence properties of the result.
Such an algorithm is equivalent to information decoding with a spin glass model and can be regarded as a simple associative memory .

We numerically simulate the Ising machine with the two operational schemes for the problem described above.
The parameter dependent performance of the GP and GC schemes shows that (i) stronger mutual coupling gives better results, and (ii) slower processes and higher pumping levels lead to higher success probabilities.
The problem size dependence of the success probability gradually decreases in the problem size up to $M = 1000$ with typical parameters. 
However, the computational time scales almost linearly with the problem size around $M = 1000$.
Also, we get a finite success probability for $M = 2000$ with realistic tuning of the parameters.

We start in Sec. \ref{sec:model} with the working equations of the laser Ising machine and explain the problems in the cubic graphs used here.
In Sec. \ref{sec:result}, we show the simulation results such as an example of dynamics of the system,
the performance dependent on parameters and the scaling of the computational time with the problem size.
Next, we discuss the properties of this machine in Sec. \ref{sec:discussion}.
Finally, we conclude the paper in Sec. \ref{sec:conclusion}.

\section{THEORETICAL MODEL}\label{sec:model}
\subsection{Langevin equations}
The field density operator of each slave laser is expanded with Glauber-Sudarshan $P(\alpha _{i})$ representation of coherent states and substituted into the master equation of
an injection-locked laser \cite{bib:laser_phys,bib:Haus_Yamamoto}.
The resulting Fokker-Planck equation corresponds to the c-number Langevin equation for the eigenvalue of the coherent state $\alpha _{i}$ via the Kramers-Moyal expansion coefficients \cite{bib:FPE}:
\begin{equation}
\frac{d}{dt} \alpha _{i} = \left[ - \frac{1}{2} \frac{\omega}{Q} + G_{i} - S_{i} |\alpha _{i}|^2 \right] \alpha _{i} + \frac{\omega}{Q} \beta _{i} + \sqrt{G_{i}} \, F_{i}. \label{eq:Langevin}
\end{equation}
Here, $\omega/Q$ is the cavity photon decay rate, $G _{i}$ is the linear gain coefficient and $S _{i}$ is the gain saturation coefficient.
$\beta _{i}$ is the sum of the amplitudes of the master injection signal and the mutual injection signal from other slave lasers.
It is assumed all the slave lasers are injection-locked to the master laser, and the Q factors of all lasers are the same.
$F_{i}$ is the Gaussian noise term and satisfy $\langle F_{i}(t) \rangle = 0$ and $\langle F_{i}(t) F_{i}(t')  \rangle = 2 \delta (t - t')$.
More rigorously, the stochastic differential equation corresponding to Eq.(\ref{eq:Langevin}) is derived with Ito's rule \cite{bib:Gardiner},
and the link between them is given by Feynman-Kac formula \cite{bib:Feynman,bib:Kac}.
Eq.(\ref{eq:Langevin}) is decomposed into the equations for the real number amplitude $\{A_{Xi}\}$ and phase $\{\phi _{Xi}\}$ for the diagonal linear polarization modes \cite{bib:Utsunomiya_Takata_Yamamoto}: 
\begin{widetext}
\begin{eqnarray}
\frac{d}{dt}A_{Xi}&=&-\frac{1}{2}\left(\frac{\omega}{Q}-E_{CVi}\right)A_{Xi}+\frac{\omega}{Q} \zeta A_M \cos \left(-\phi_{Xi}\right)
- \sum_{j \neq i}\frac{1}{2} \frac{\omega}{Q} \xi_{ij} \left\{ A_{Xj} \cos \big[\phi_{Xj}-\phi_{Xi} \big] - A_{\bar{X}j} \cos \big[\phi_{\bar{X}j}-\phi_{Xi} \big] \right\} + F_{AXi}, \label{eq:SDE AXi}\\
\frac{d}{dt}\phi_{Xi}& = & \frac{\omega}{Q}\frac{1}{A_{Xi}} \Big\{ \zeta A_M \sin \left(-\phi_{Xi}\right)
- \sum_{j\neq i}\frac{1}{2}\xi_{ij}\left[ A_{Xj} \sin \big(\phi_{Xj} - \phi_{Xi} \big)- A_{\bar{X}j} \sin \big(\phi_{\bar{X}j} - \phi_{Xi} \big)\right] \Big\}
+ F_{\phi Xi}, \label{eq:SDE phiXi}
\end{eqnarray}
\end{widetext}
where $\{ X, \bar{X} \} = \{D, \bar{D}\}$ are combinations of the two linear polarization modes along $\pm 45^\circ$ with 
respect to the vertical linear polarization.  $i, j = 1,2, \ldots M$ are indices for slave lasers and $M$ is the problem size (the number of Ising spins).
$\zeta$ is the coupling coefficient for the master laser signal, whose amplitude is denoted by $A_M$. 
The mutual coupling constant $\xi _{ij} = \alpha(t) \, J_{ij}$ is determined by the transmission coefficient $\alpha(t)$ and the Ising interaction parameter $J_{ij}$.
The gain term $G_{i} - S_{i} |\alpha _{i}|^2$ and the diffusion coefficient $G _{i}$ in Eq.(\ref{eq:Langevin}) correspond to the gain $E_{CVi}$ induced by the active carrier
and the pumping rate $P(t)$, respectively.
We add the following equation of motion for the carrier number $\{N_{Ci}\}$
\begin{equation}
\frac{d}{dt} N_{Ci} = P(t) - \frac{N_{Ci}}{\tau _{sp}} - E_{CVi} (A_{Di}^2 + A_{\bar{D}i}^2) + F_{NCi}, \label{eq:rateNC}
\end{equation}
where $E_{CVi} = \beta N_{Ci}/\tau _{sp}$ and $\beta$ is the spontaneous emission coupling efficiency.
$F_{AXi}$, $F_{\phi Xi}$ and $F_{NCi}$ are noise terms.
We consider only spontaneous emission noise with the rate $E_{CVi}$ in the numerical simulation,
since it is a dominant noise source fluctuating the phases of the diagonal polarization modes \cite{bib:Utsunomiya_Takata_Yamamoto,bib:Henry}.
Spontaneous emission processes are treated as the discrete random Poisson process for each numerical integration time step $\Delta t$,
and each spontaneous emission photon coupled to the laser field has unit norm and random phase.
Ref. \onlinecite{bib:Henry} has shown that this model reproduces the correct Langevin forces in the continuous time unit.
We numerically integrate Eqs. (\ref{eq:SDE AXi}), (\ref{eq:SDE phiXi}) and (\ref{eq:rateNC}) and continuously monitor the mapped collective spins $\sigma _i = (A_{Ri} - A_{Li})/\sqrt{A_{Ri}^2 + A_{Li}^2}$
using Eqs.(35) and (36) in Ref. \onlinecite{bib:Utsunomiya_Takata_Yamamoto}.
Here $A_{Ri}$ and $A_{Li}$ are the slowly-varying amplitudes of the right and left circularly polarized modes of the $i$th slave laser.
Signs of $\{\sigma _i\}$ are used to determine the Ising spin $\{\sigma _i\} = +1$ or $-1$.

\subsection{Problem setting}
\begin{figure}[htbp]
	\includegraphics[clip, scale=0.25]{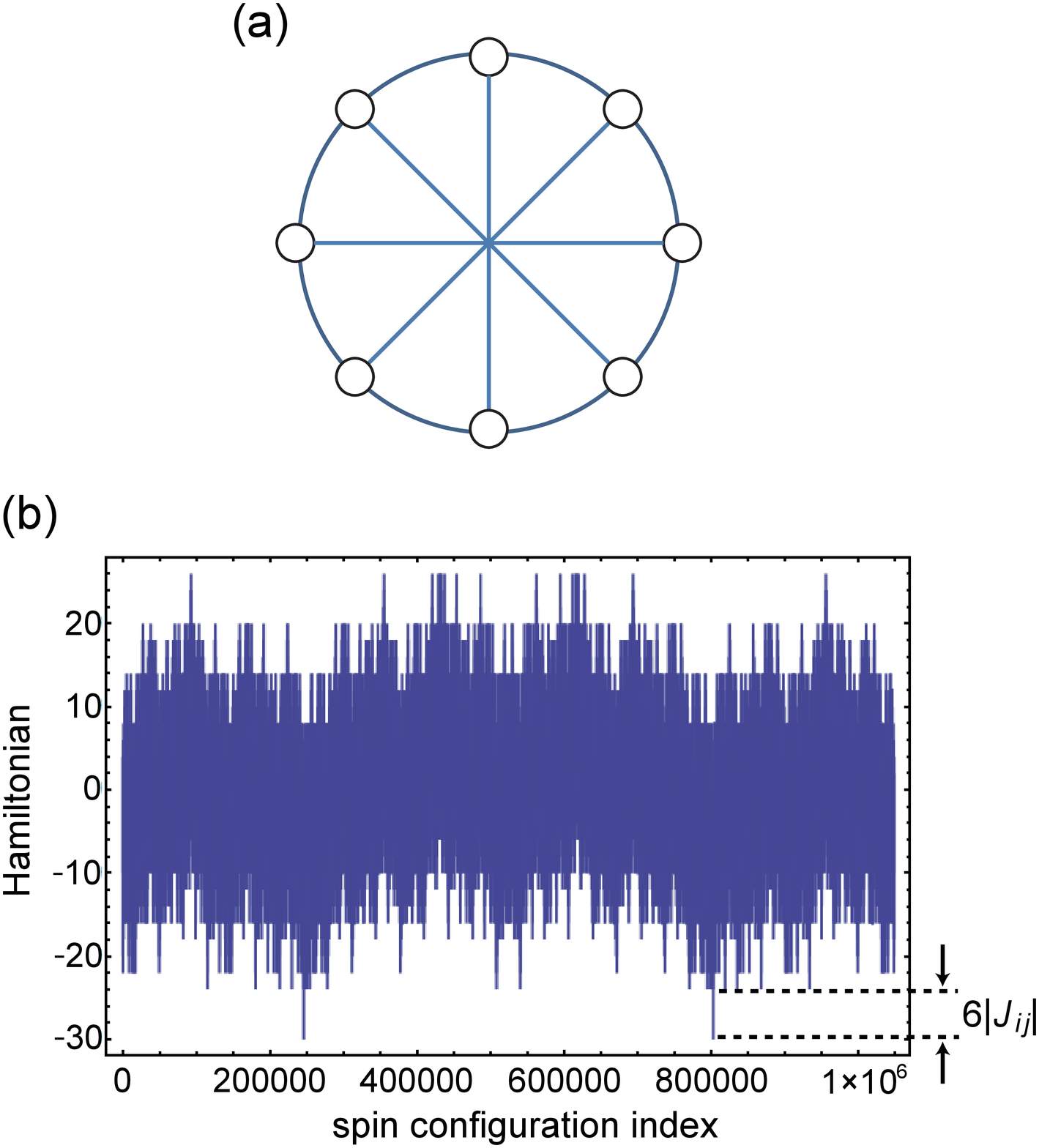}
	\caption{(a) An example of cubic graphs considered here with $M=8$. (b) An example of energy (or effective loss) landscape in a problem with $M=20$.}\label{fig:prob}
\end{figure}

In Fig. \ref{fig:prob}(a), we show an example of cubic graphs with $M = 8$ for the data search problem considered in this study.
Here, the nodes align in a ring shape. Each node has the couplings to the nearest neighbors and the diameter chord.
The Ising coupling term are set for a picked target state $\{\sigma _{Ai}\} = \{+1, -1\}$ as
\begin{equation}
J_{ij} = \begin{cases}
					-\sigma _{Ai}\sigma _{Aj} & ({\rm if \ node } \ i \ {\rm and } \ j \ {\rm are \ connected}), \\
					0 & ({\rm otherwise}).
			 \end{cases} \label{eq:Jij}
\end{equation}
Fig. \ref{fig:prob}(b) is an example of energy (net loss) landscapes in this Ising problem with $M = 20$.
This Hamiltonian does not have any frustration so that it has only two ground states, while the total number of states is $10^6$.
By definition of the couplings, one of the ground states is identical to the target state.
The difference between the mapped energy of the ground states ($E_{\rm g}$) and the first excited states ($E_{\rm 1e}$) is independent of $M$ and equal to $6 |J_{ij}| = 6$.
This is because of the highly symmetric property of the graph. We show a concise proof with induction in the Appendix.
Similar problems are also discussed in the context of information coding and decoding with statistical mechanics \cite{bib:Nishimori}.

\section{SIMULATION RESULT}\label{sec:result}
\subsection{Dynamics of the system}
Here, we show a variety of simulation results of the machine for the setting described in the previous sections.
\begin{figure*}[htbp]
	\includegraphics[clip, trim= 0 0 0 0, scale=0.25]{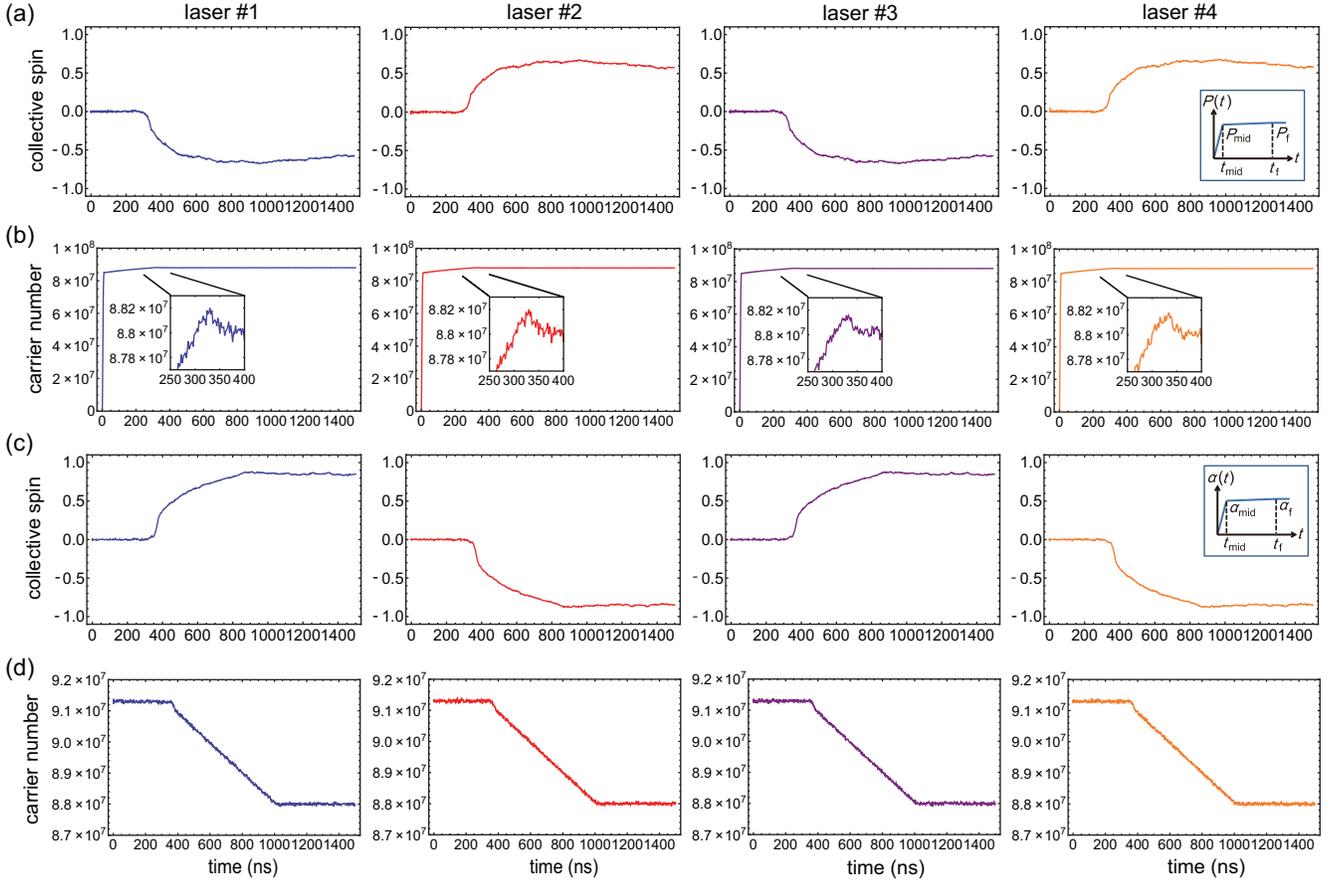}
	\caption{A set of examples of time evolution of (a) the polarization states (collective spins) and (b) the carrier numbers for a target file with $M=4$ in the gradual pumping scheme. Transit of (c) the spins and (d) the carrier numbers in the gradual coupling scheme.
	The inset of (a) and (c) are the scheduling of the pumping level and coupling coefficient.}\label{fig:transit}
\end{figure*}
Fig. \ref{fig:transit}(a) and (b) show the time evolution of the polarization states (collective spins) and the carrier numbers in the GP scheme for $M=4$.
The inset of Fig. \ref{fig:transit}(a) is the schematic pumping schedule of all slave lasers. The pumping schedule is composed of the two linear parts
(rapid and slow increase) with the turning point $t_{\rm mid} = 10 \ {\rm ns}$ (fixed for all the simulations here),
and the pumping gets constant at $t = t_{\rm f}$.
The process time is defined as $t_{\rm P} \equiv t_{\rm f} - t_{\rm mid}$.
The parameters commonly used in this study are $\omega/Q = 10^{11} \ {\rm s}^{-1}$, $\tau _{\rm sp} = 10^{-9} \ {\rm s}$ and $\beta = 10^{-6}$.
The threshold pumping current is $I_{\rm th} = e P_{\rm th} = 16 \ {\rm mA}$, where $e$ is the elementary charge and $P_{\rm th}$ is the threshold pumping rate.

In the GP scheme, the master laser signal is injected at $t = 0$ and the slave laser amplifies this signal so that the collective spin
is prepared in $\sigma _{i} = 0$ for all $i$.
With a proper intermediate value $P_{\rm mid} = P(t_{\rm mid})$, the collective spins bifurcate at a time between $t_{\rm mid}$ and $t_{\rm f}$ (here $t_{\rm f} = 1010 \ {\rm ns}$).
The plot of carrier numbers has a small overshoot at the bifurcation point, which indicates that once the minimum loss (ground) state is selected, the gain is decreased accordingly.
Fig. \ref{fig:transit}(c) and (d) show the transient examples of the collective spins and carrier numbers in the GC scheme,
and the inset of Fig. \ref{fig:transit}(c) is the coupling schedule. In this scheme, the total gain monotonically decreases as the optical coupling increases.

Here, all the slave lasers bifurcate their polarizations and produce the clear signals in both schemes.
However, in a large problem size the polarization bifurcation can be degraded by noise and the system can fail to find a ground state.
In this case, the values of the collective spins get non-uniform and some slave lasers have nearly $\sigma _i \sim 0$.
Also, the system occasionally shows strong bifurcation into $D$ and $\bar{D}$ polarization modes (not $R$ and $L$ modes).
In this case, however, a correct ground state can be recovered with the decision based on the $\{D, \bar{D}\}$
basis, i.e. via the new definition of the collective spin $\sigma _i = (A_{Di} - A_{\bar{D}i})/\sqrt{A_{Di}^2 + A_{\bar{D}i}^2}$.
Note that the coupling optics with a horizontal polarizer \cite{bib:Utsunomiya_Takata_Yamamoto,bib:Takata_Utsunomiya_Yamamoto} can implement the identical
loss modulation for $D$ and $\bar{D}$ modes as the $R$ and $L$ modes.

\subsection{Parameter dependence of performance}
\begin{figure}[htbp]
	\includegraphics[clip, scale=0.19]{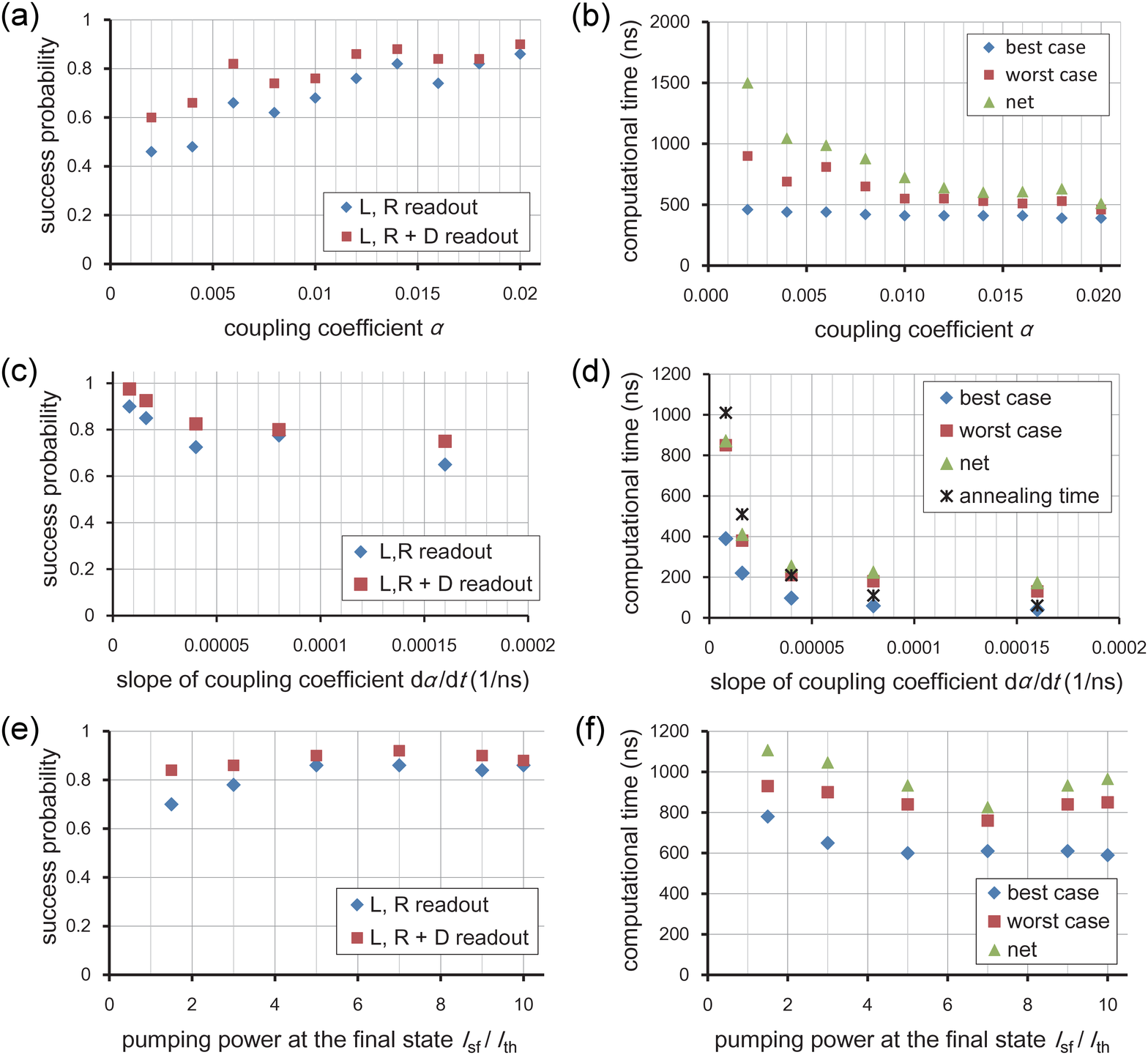}
	\caption{(a) The success probability and (b) the net computational time dependent on the coupling coefficient $\alpha$ in the GP scheme. 
	(c) The success probability and (d) the net computational time dependent on the process speed ${\rm d}\alpha /{\rm d}t$ in the GC scheme.
	(e) The success probability and (f) the net computational time dependent on the normalized final pumping level $I_{\rm sf}/I_{\rm th}$ in the GP scheme.}\label{fig:parameter dep}
\end{figure}

Fig. \ref{fig:parameter dep}(a) and (b) show the dependence of the success probability and computational time on the coupling coefficient $\alpha$ in the GP scheme.
We pick a 100-spin problem and run the simulation 50 times for each value of the parameter considered.
We change the coupling coefficient $\zeta$ of the master laser signal proportional to $\alpha$ to keep the good balance between them. Other parameters are fixed.
We define the computational time as the time where a ground state is found with all the collective spins above a certain threshold $|\sigma _{\rm i}| = 0.071$.
This value achieves a measurement signal-to-noise ratio of $ S/N \sim 10^3$ with a detection quantum efficiency $\eta _{D} = 0.01$, an integration time $T = 1 \ {\rm ns}$ and a total photon number $n_{Ti} = 10^4$ in a slave laser \cite{bib:Utsunomiya_Takata_Yamamoto}.
Here, $n_{Ti} \ll 1/\beta = 10^6$, thus any pumping rates above the threshold have better $S/N$ ratios.
The resulting upper bound of the measurement error rate is $P_{\rm e} \sim 6.4 \times 10^{-57}$ for a single slave laser and negligible even in a 1000 laser system.
Collective spins are basically monitored with the circular polarization basis.
The check with the diagonal mode basis is conducted at the final state and reflected in the success probability.
The net computational time denotes the worst computational time divided by the success probability. 
Fig. \ref{fig:parameter dep}(a) shows that a stronger mutual coupling $\alpha$ tends to give a higher success probability.
Also, a relatively large noise effect with a small $\alpha$ leads to bifurcation in the $D$ or $\bar{D}$ mode then a low overall success probability.
In Fig. \ref{fig:parameter dep}(b), the computational time decreases with increasing $\alpha$ and saturates at $\alpha \gtrsim 0.01$.
With a small $\alpha$, the system can give a slow transition process and need a long time to find a ground state.

Fig. \ref{fig:parameter dep}(c) and (d) show the dependence of the success probability and simulated computational time on the rate of increasing the mutual coupling between
$t_{\rm mid}$ and $t_{\rm f}$ in the GC scheme.
Fig. \ref{fig:parameter dep}(c) shows a clear tendency that a slower process gives a better success probability.
The process with ${\rm d}\alpha/{\rm d}t \rightarrow \infty$ can be regarded as the abrupt introduction of the mutual injection in the previous study,
thus it can be said the GC scheme improves the success probability compared to the abrupt scheme \cite{bib:Utsunomiya_Takata_Yamamoto,bib:Takata_Utsunomiya_Yamamoto}.
We have got a similar result also in the GP scheme.
In Fig. \ref{fig:parameter dep}(d), a slower process (a longer $t_{\rm P}$) requires a longer computational time nearly linear with $t_{\rm P}$ when other conditions keep unchanged.

Fig. \ref{fig:parameter dep}(e) and (f) show the dependence of the success probability and computational time on the final pumping level $I_{\rm sf}$ normalized by the threshold pumping $I_{\rm th}$ in the GP scheme.
Here, the end time of the process $t_{\rm f}$ is fixed, thus the slope ${\rm d}P/{\rm d}t$ is variable.
A large coherent slave laser signal with a high pumping level leads to clear computation against noise.
At the same time, however, a higher pumping results in a higher slope ${\rm d}P/{\rm d}t$ which degrades the performance as seen in Fig. \ref{fig:parameter dep}(c) and (d).
Thus, Fig. \ref{fig:parameter dep}(e) and (f) have the optimum points at $I_{\rm sf}/I_{\rm th} = 7$ out of these two effects.

\subsection{Net computational time}
\begin{figure}[htbp]
	\includegraphics[clip,trim= 0 0 0 0, scale=0.33]{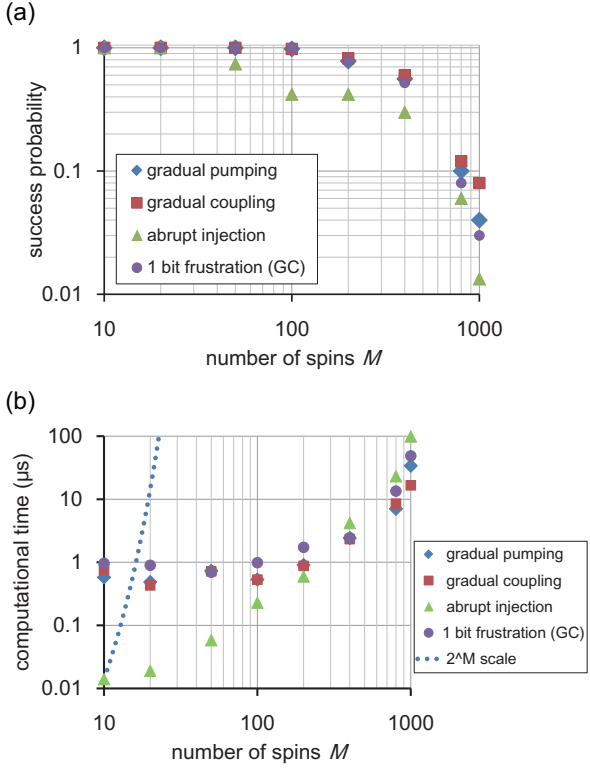}
	\caption{(a) The success probability and (b) the net computational time dependent on the number of spins $M$.}\label{fig:M dep}
\end{figure}

Fig. \ref{fig:M dep}(a) and (b) show the problem size dependence of the success probability and the net computational time for the three schemes.
For each problem size, we pick five targets and run the simulation ten times for each target (for $M=1000$ in the abrupt scheme, we take 15 targets).
The success probability is determined with the combined readouts using both circular and diagonal measurement bases.
The ratio $P_{\rm mid}/P_{\rm f} = 0.5$ (GP) and $\alpha _{\rm mid}/\alpha _{\rm f} = 0.6$ (GC) are selected to obtain close bifurcation times.
Other parameters are identical: $\alpha({\rm GP}) = \alpha _{\rm f}({\rm GC}) = 0.02$, $I _{\rm sf}({\rm GP}) = I _{\rm s}({\rm GC}) = 3 \ I _{\rm th}$ and $t_{\rm P} = 1 \ {\rm \mu s}$.
In Fig. \ref{fig:M dep}(a), the success probability shows the gradual decrease in large problem sizes in the gradual pumping and coupling schemes.
In contrast, the probability for the abrupt scheme gets lower.
In Fig. \ref{fig:M dep}(b), the net computational time for the GP and GC schemes keep nearly constant up to $M = 200$ and turn into nearly linear increase.
On the other hand, the computational time for the abrupt scheme monotonically increases and gets longer than those for the gradual pumping and coupling schemes over $M=400$.
Here, we add the performance of the GC scheme with a sign flip for a random coupling coefficient $J_{ij} \ (=J_{ji})$.
This keeps a good success probability and a better computational time than those for the abrupt scheme over $M=400$.
A sign flip of $J_{ij}$ induces a single frustrated part in the original ground state.
However, originally $E_{\rm g} - E_{\rm 1e} = 6$, thus the ground states remain still.
This means the cubic graph system is insusceptible to a single phase flip error in the mutual coupling configuration,
i.e. the encoded data of the target state.
Note that a simpler implementation with 1D rings does not have such robustness.
We add the scaling of brute force search $(2^{M})$ for the information.
The net computational time with the laser Ising machine has a much better scaling,
probably because of the simple structure of the benchmarked problem.

We can increase the success probability for a large problem size by using even an slower process and a higher pumping power.
We have got the success probability of $34\%$ with $ \{ t_{\rm P}, I_{\rm sf}/I_{\rm th} \} = \{3 \ {\rm \mu s}, 20\} $ for $M = 1000$ and 
$15\%$ with $ \{ t_{\rm P}, I_{\rm sf}/I_{\rm th} \} = \{10 \ {\rm  \mu s}, 50\} $  for $M = 2000$ in the GC scheme.
Here the numbers of the candidate states $2^M$ differ by $2^{1000} \simeq 10^{300}$ times between $M = 1000$ and 2000,
but the system does not need a large ratio for $t_{\rm P}$ and $I_{\rm sf}$ to obtain a reasonable success probability.
This indicates the time and pumping resource required to solve this problem does not grow exponentially even for a relatively large problem size.
Here, we note it has been shown some open quantum systems with frustration-free Hamiltonians can converge at steady states in sub-exponential time \cite{bib:Verstraete}.

\section{DISCUSSION}\label{sec:discussion}
\subsection{Success probability}
\begin{figure}[htbp]
	\includegraphics[clip,trim= 0 34 0 10, scale=0.33]{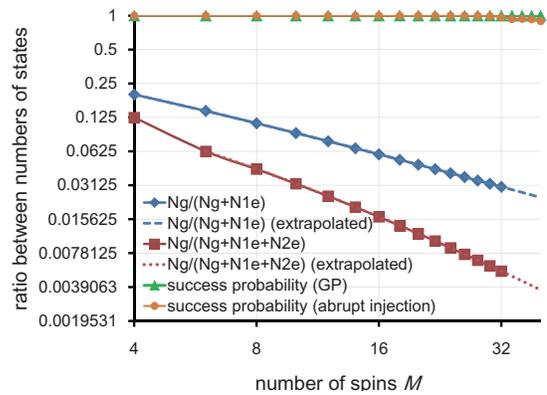}
	\caption{The ratio of the number of ground states in a single problem up to $M = 40$. The success probability with the GP scheme and the abrupt scheme are concurrently shown.}\label{fig:prob_small_size}
\end{figure}

Here we discuss properties of the coherent laser machine from various points of view.
First, we take a closer look at the success probability. Fig. \ref{fig:prob_small_size} shows the ratio of the number of degenerate ground states $N_{\rm g}$ to those of degenerate excited states.
Here, $N_{\rm 1e}$ and $N_{\rm 2e}$ denote the number of first and second excited states, respectively.
These values have been computed with the brute force search up to $M = 32$, and extrapolated for larger problem sizes.
We add the success probability with the GP scheme out of 50 trials and that with the abrupt scheme out of 200 trials for each $M$.
We took only a single problem randomly for each $M$, however, the computed numbers of states are exactly along with the extrapolation curves
$N_{\rm 1e} = 2 M$ and $N_{\rm 2e} = M (M + 6)/4$ for $M \ge 6$.
This fact indicates the universality of the considered problem due to its graphic symmetry.
Here, from the data of failure, we expect the local minima which mostly affects the laser machine are second excited states.
The ratios $N_{\rm g}/(N_{\rm g} + N_{\rm 1e})$ and $N_{\rm g}/(N_{\rm g} + N_{\rm 1e} + N_{\rm 2e})$ are inversely proportional to $M$ and $M^2$,
while the laser network keeps the high success probability: 100 \% for the GP scheme and $\ge 90$ \% for the abrupt pumping scheme up to $M=40$.
This clearly shows this machine is distinct from simple probabilistic searches and preferably takes the minimum gain states, especially in the gradual schemes.

A possible reason for failure in the GP scheme can be probabilistic errors due to a finite time necessary for the system to detect and amplify the final state.
The machine keeps increasing the gain in the process because of the unknown gain of the ground states.
Therefore, if it remains at the initial state long, it may get an enough gain with which an excited state can oscillate.
This malfunction can be overcome by slowing the pumping process.

Another possibility is disturbance by noise.
The sum of the gain coefficients at the steady state, which determines the absolute photonic loss of the whole system, is given by \cite{bib:Utsunomiya_Takata_Yamamoto}
\begin{equation}
\sum_{i} E_{CVi} = M \frac{\omega}{Q} - \zeta \sum_{i} \sqrt{2 - \sigma_{i}^{2}} + \alpha \sum_{i<j} J_{ij} \sigma_{i} \sigma_{j},
\end{equation}
where the first, second and third term are associated with the cold cavity dissipation, the master and mutual injection, respectively.
Thus, when the mutual injection gets dominant at the phase transition, a minimum loss state corresponds to a ground state of Ising Hamiltonian.
Then, the energy gap between a ground state and a first excited state is expected to depend only on the coupling coefficient $\alpha J_{ij}$.
On the other hand, the total noise power in the whole system is proportional to $M$.
Thus, the success probability is considered to decrease with the problem size $\propto M^{-1}$ above a certain threshold,
and the simulation result seems to support this behavior.
Note that the GC scheme also shows such characteristics.
The difference in performance of the GP and GC methods is not significant and can be explained by the small difference in the parameter ratio
($P_{\rm mid}/P_{\rm f}$ and $\alpha _{\rm mid}/\alpha _{\rm f}$) and then the slope of these.
Thus, we expect these two have almost the same computational ability.

\subsection{Response speed of the machine}
Next, we discuss the response of the laser network machine at the onset of bifurcation.
If we neglect the noise terms including spontaneous emission, the equation for the amplitude difference in circular polarization is given by \cite{bib:Utsunomiya_Takata_Yamamoto,bib:Takata_Utsunomiya_Yamamoto}
\begin{widetext}
	\begin{equation}
		\frac{d}{dt}\left(A_{Ri} - A_{Li}\right) = - \frac{1}{2} \left( \frac{\omega}{Q} - E_{CVi}\right) \left(A_{Ri} - A_{Li}\right) - \frac{\omega}{Q} \sum_{j \neq i} \xi_{ij} \left(A_{Rj} - A_{Lj}\right).\label{eq:difference}
	\end{equation}
\end{widetext}
Here, note that the master signal term vanishes in the equation.
We do not consider the time dependence of $E_{CVi}$ i.e. $N_{Ci}$ because the field response is much faster: $\omega/Q \gg 1/\tau_{\rm sp}$.
By assuming that all the slave lasers are homogeneously driven into a target state as shown in Fig. \ref{fig:transit}, we can set $\xi _{ij} (A_{Rj} - A_{Lj}) \approx - \alpha |A_{Ri} - A_{Li}|$.
With this and Eq. (\ref{eq:difference}), we derive the approximate response around the bifurcation point by the cubic graph system as
\begin{equation}
|\sigma _i | = \frac{|A_{Ri} - A_{Li}|}{\sqrt{n_{Ti}}} \approx C \exp \biggl[ \biggl( \frac{- \omega /Q + E_{CVi}}{2} + 3 \alpha \frac{\omega}{Q} \biggr) t \biggr], \label{eq:approx}
\end{equation}
where the small change in $n_{Ti}$ due to injection is neglected. $C$ is an integral constant.
We see, when the bifurcation occurs, the polarization configuration is formed with the exponentially fast modulation by the injection signal in ideal cases.
The gradual schemes enhance the success probability by changing in $E_{CVi}$ or $\alpha$ slowly.
Also, Eq. (\ref{eq:approx}) indicates the magnitude of injection signals directly affect the response speed of the system \cite{bib:Takata_Utsunomiya_Yamamoto}, and possibly the distribution of the values of collective spins.
For NP-hard problems with frustration, the system often has imbalance of the response speeds and the collective spins in the slave lasers then get unable to read a correct answer.
Thus, the previous abrupt injection scheme needs a self-learning algorithm to compensate for it \cite{bib:Kai}.

\subsection{Algorithmic properties}
Finally, we discuss the laser Ising machine as an algorithm.
The laser machine itself is a kind of relaxation algorithms in the meaning that it maps discrete variables to continuous variables.
This point is in common with the well-known semi-definite programming (SDP) algorithm for optimization problems \cite{bib:Vazirani}.
The laser network machine utilizes the three-dimensional Poincar\'e sphere for each spin, while SDP exploits an $M$-dimensional vector.
SDP requires an algorithm to drive the state to an optimum one, as typified by interior point methods,
and such a method costs the order ${\rm O}(M^3)$ of basic operations at least \cite{bib:Korte}.
Also, if SDP can be solved with an arbitrarily small error,
a rounding algorithm to map the vectors back to discrete variables guarantees the worst approximation rate of 87.8$\%$ for the MAX-CUT problem \cite{bib:0.878_alg}.
The laser network depends on the dynamics of the coupled differential equations for finding an optimum state.
As an experimental system, the laser machine shows a nearly ${\rm O}(M)$ scaling of the computational time up to $M=1000$ for the data search problem studied here.
Also, Ref. \onlinecite{bib:Kai} shows some good results of comparison between the simulated performance of the laser machine and SDP for hard optimization problems.
As an computational simulation, however, we need some additional time complexities.
For limited numbers of couplings for each spin (sparse interaction matrices), numerical integration costs ${\rm O}(M)$ operations multiplied by the order of the net computational time the machine needs.
Thus, the overall time complexity for simulations of the laser machine here is about ${\rm O}(M^2)$.
In addition, the numerical integration method used here (fixed-step $4 \times 4$ Runge-Kutta method and additional stochastic noise) requires a large coefficient outside the complexity.
Thus the simulation for the laser network seems to take more time than SDP in practice.

\section{CONCLUSION}\label{sec:conclusion}
In conclusion, we have studied two gradual driving schemes for the coherent Ising machine.
The GP scheme is to find a minimum-loss ground state by increasing the gain (or increasing the temperature in the negative region)
and the GC scheme to exploit the gradual path between the two minimum loss ground states of the initial and final mapped Hamiltonian.
The numerical simulation with the Langevin equations on a data search problem with the Ising model in particular cubic graphs shows that
we can improve the success probability with slowing the pumping or coupling schedule, and increasing the final pumping power.
With typical parameters, the laser network and these gradual schemes give an almost constant computational time up to $M = 200$, and turns into a nearly linear
scale holding up to $M=1000$ in the problem.
This scaling is better than that with the previously studied abrupt scheme in large problem sizes.
In addition, simulations with varying parameters indicates that the required time and pumping resources to find a target do not scale exponentially up to $M = 2000$.
Now we are aiming at implementing an experimental system that works much faster and hopefully better than algorithms for digital computers,
especially on hard optimization problems.

\begin{acknowledgments}
We thank Kai Wen for his preliminary work and discussion and Ken-ichi Kawarabayashi for fruitful discussion about approximate optimization algorithms.
This work is supported by the JSPS through its FIRST program, Navy/SPAWAR Grant N 66001-09-1-2024, and the Special Coordination Funds
for Promoting Science and Technology. K.T. is a JSPS research fellow and thanks for Grant-in-Aid for JSPS Fellows.
\end{acknowledgments}

\appendix*
\section{Energy of degenerate first excited states}
\begin{figure}[htbp]
	\begin{center}
		\includegraphics[clip, trim = 0 0 0 0, scale=0.30]{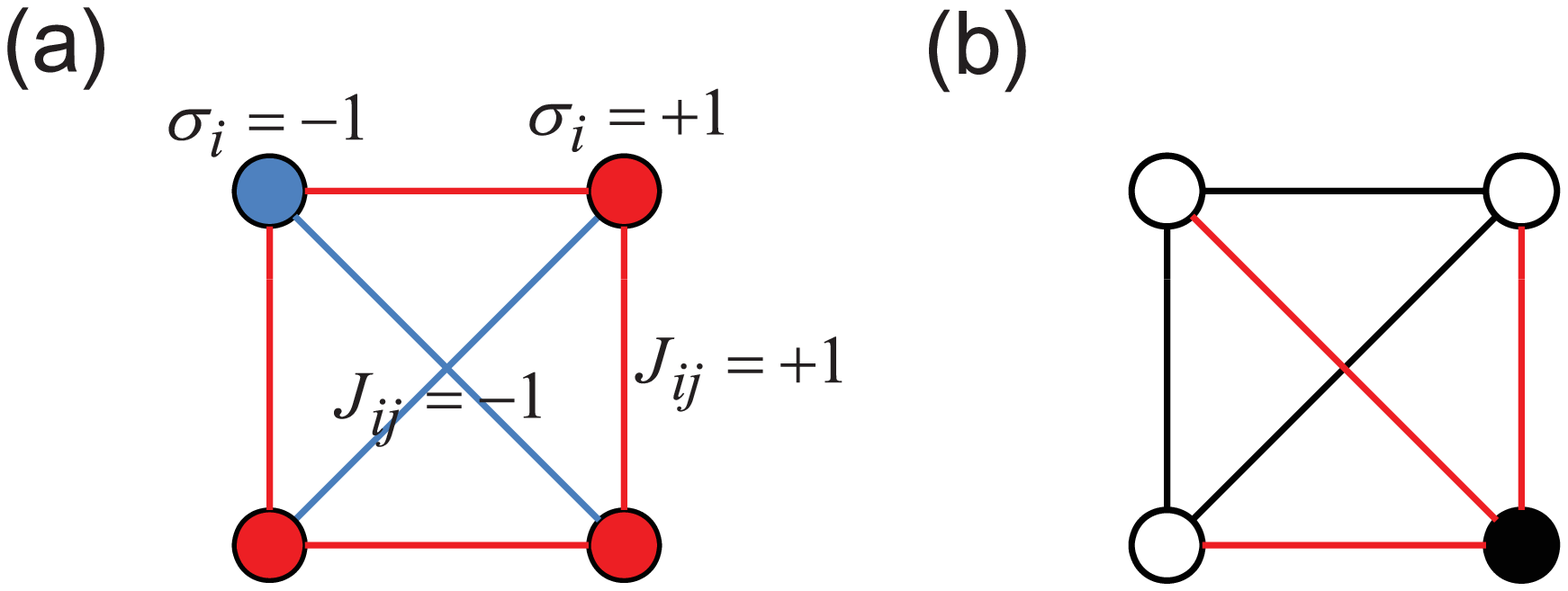}
		\caption{(a) Schematic illustration for the Ising spins and coupling between them in the graph of $M=4$. (b) The converted graph corresponding to (a).}\label{fig:4spins}
	\end{center}
	\begin{center}
		\includegraphics[clip, trim = 0 0 0 50, scale=0.30]{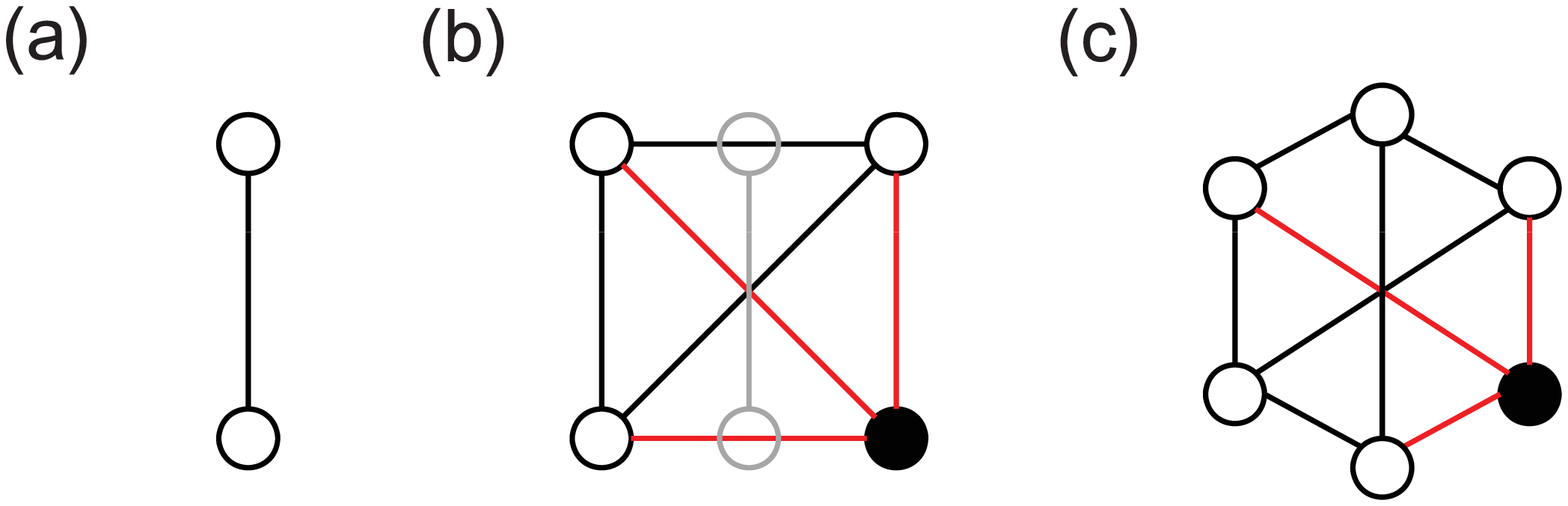}
		\caption{(a) The partial graph added in order to make the graph with $M=2l+2$ from that with $M=2l$. (b) The way of adding the partial grpah. (c) The result of two-node addition for the example in (b).}\label{fig:addition}
	\end{center}
\end{figure}

Here, we prove that the energy of degenerate first excited states is $E_g + 6$ ($E_g$: the energy of ground states) for the problem of
the symmetric cubic graphs without frustration, which is considered in our study. The most fundamental example with $M = 4$ is shown in Fig. \ref{fig:4spins}.
The problem used here constitutes the spins and Ising interaction, both of which are valued $+1$ or $-1$, as in Fig. \ref{fig:4spins}(a).
Here, we reinterpret this in terms of the stability of couplings as shown in Fig. \ref{fig:4spins}(b).
In Fig. \ref{fig:4spins}(b), nodes of the same color (e.g. white and white) have the stable coupling expressed with black edges between them,
and each of them decreases the energy function by unity.
Also, those of different colors (white and black) have the unstable coupling colored red, increasing the energy by the same amount.
Thus, flipping one spin changes the energy by two for every coupling.
With this procedure, we can consider all the instances and states simply by counting the number of edges indicating the unstable coupling.
We note that the graph with $E_g + 6$ means it has three unstable edges.

We use induction to prove that the first excited states have $E_g + 6$.
For $M = 4$, we can show this by checking all the 16 patterns possible to be reached by flipping spins (changing colors of nodes) from the
ground state composed of, say, all white nodes. Also, we assume that this holds for $M = 2l$, where $l$ is an integer larger than one.
Here, we prove that this is valid also for $M=2l+2$.

\begin{figure}[htbp]
	\begin{center}
		\includegraphics[clip, trim = 0 0 0 0, scale=0.30]{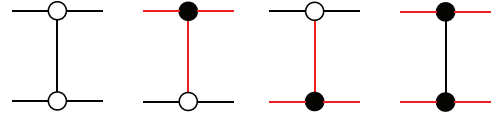}
		\caption{Possible node and edge configuration of the additional part in the first case.}\label{fig:case1}
	\end{center}
\end{figure}

\begin{figure}[htbp]
	\begin{center}
		\includegraphics[clip, trim = 0 0 0 0, scale=0.30]{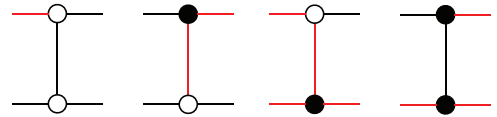}
		\caption{Possible node and edge configuration of the additional part in the second case.}\label{fig:case2}
	\end{center}
\end{figure}

\begin{figure}[htbp]
	\begin{center}
		\includegraphics[clip, trim = 0 0 0 0, scale=0.30]{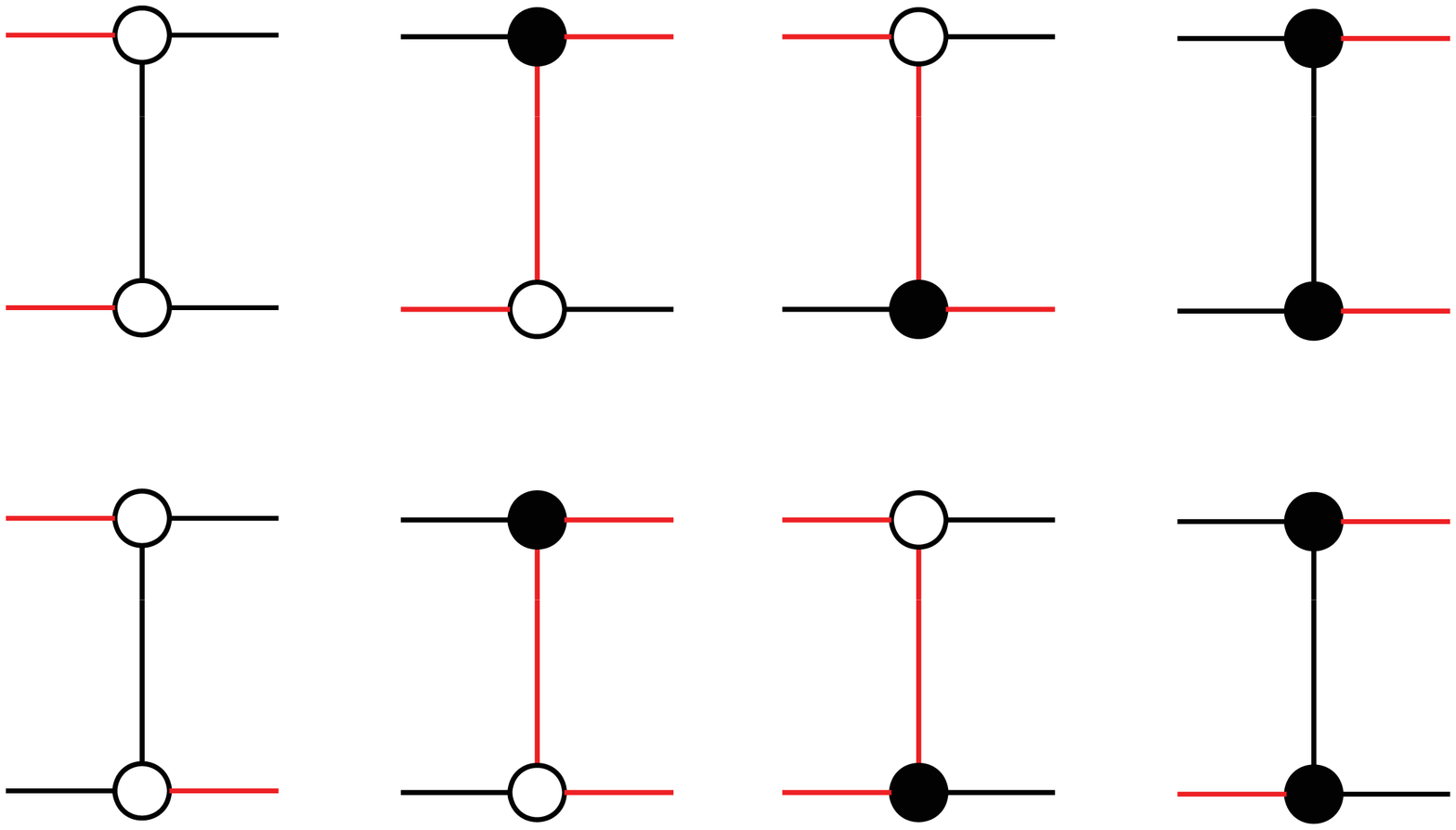}
		\caption{Possible node and edge configuration of the additional part in the third case.}\label{fig:case3}
	\end{center}
\end{figure}

We can make the graph with $M=2l+2$ by adding the two-node partial graph shown in Fig. \ref{fig:addition}(a) to that with $M = 2l$.
Fig. \ref{fig:addition}(b) shows the way of adding the partial graph with an example of $M=4$. The result is shown in Fig. \ref{fig:addition}(c).
Here, each edge piercing an additional node in Fig. \ref{fig:addition}(b) is divided into two edges, and the couplings for these edges are affected by the additional nodes.
Thanks to the assumption for $M = 2l$, it is enough to consider the change induced by the additional part.
Also, note that it does not matter where the additional part comes, due to the symmetry of the graph.
From the viewpoints of positions of the unstable edges in the graph with $M = 2l$, it is sufficient to examine the three cases as follows.
\begin{enumerate}
\setlength{\labelsep}{2em}
\item No unstable edges divided by the additional nodes, and 0, 3 or more unstable edges outside the additional graph
\end{enumerate}
Fig. \ref{fig:case1} shows the possible node configuration in this case. Here, the number of unstable edges is 0, 3 or 4.
This means the total number of unstable edges is 0, 3 or more also for the graph with $M = 2l+2$, and the possible energy the system can take
is $E_g$, $E_g + 6$ or more.
\begin{enumerate}
\setlength{\labelsep}{2em}
\setcounter{enumi}{1}
\item 1 unstable edge divided by an additional node, and 2 or more unstable edges outside the additional graph
\end{enumerate}
Due to the symmetry of the graph, it is enough to consider the four patterns in Fig. \ref{fig:case2} here.
There are one or more unstable edges in the additional part, thus the total number of unstable coupling is 3 or more for $M = 2l+2$.
\begin{enumerate}
\setlength{\labelsep}{2em}
\setcounter{enumi}{2}
\item 2 unstable edges divided by an additional nodes, and 1 or more unstable edges outside the additional graph
\end{enumerate}
According to configuration of the nodes connected to the additional part, we consider the eight patterns shown in Fig. \ref{fig:case3} in this case.
We can see that two or more unstable edges exist in the any patterns of Fig. \ref{fig:case3}, thus the possible energy for this case in $M = 2l+2$ is $E_g + 6$ or more.

From all the cases above, the energy of the first excited states for $M = 2l+2$ is also $E_g + 6$.
Here, note that the first case clearly shows the existence of the states with the energy that is exactly $E_g + 6$.
Therefore, the proposition is proven for all $M = 2l \,\, (l=2,3,...)$ with inductive application of the discussion above.

%


\end{document}